%% file: main.tex
\ifx\synctex\undefined\else\synctex=1\fi
\pdfoutput=1
\documentclass{eptcs}

\providecommand{\event}{VPT 2015}

\usepackage{graphicx}
\usepackage{amsmath} 
\usepackage{algorithm2e}
\usepackage{hyperref}
\usepackage{amssymb}
\usepackage{upquote}
\usepackage{fancyvrb}

\usepackage{wrapfig}
 
\usepackage[draft]{commenting}

\declareauthor{jg}{John}{blue}
\declareauthor{pg}{Pierre}{red!50}
\declareauthor{bk}{Bisho}{orange!50}

\title{ Decomposition by tree dimension in Horn clause verification \thanks{The research leading to these results
   has been supported by the EU
  FP7 project 318337, \emph{ENTRA - Whole-Systems Energy Transparency},  the EU FP7 project 611004, \emph{coordination and support action ICT-Energy} and
   Danish Research Council grant FNU-10-084290.}}

\author{Bishoksan Kafle \institute{Roskilde University, Denmark} \email{kafle@ruc.dk}
\and
John P. Gallagher \institute{Roskilde University, Denmark} 
\institute{IMDEA Software Institute, Spain}
 \email{jpg@ruc.dk}
\and 
 Pierre  Ganty \institute{IMDEA Software Institute, Spain} \email{pierre.ganty@imdea.org}
 }

\def\titlerunning{Decomposition by tree dimension in Horn clause verification}
\def\authorrunning{ Kafle, Gallagher and  Ganty}

\begin{document}
\maketitle

\input{localdefs}

\begin{abstract}

In this paper we investigate the use of the concept of \emph{tree
dimension} in Horn clause analysis and verification.  The dimension of a
tree is a measure of its non-linearity -- for example a list of any
length has dimension zero while a complete binary tree has dimension
equal to its height. We apply this concept to trees corresponding to
Horn clause derivations. A given set of Horn clauses $P$ can be
transformed into a new set of clauses $P^{\atmost{k}}$, whose derivation
trees are the subset of $P$'s derivation trees with dimension at most
$k$. Similarly, a set of clauses $P^{\exceeds{k}}$ can be obtained from
$P$ whose derivation trees have dimension at least $k+1$. In order to
prove some property of all derivations of $P$, we systematically apply
these transformations, for various values of $k$, to decompose the proof
into separate proofs for $P^{\atmost{k}}$ and $P^{\exceeds{k}}$ (which could be executed in
parallel). We show some preliminary results indicating that
decomposition by tree dimension is a potentially useful proof technique.
We also investigate the use of existing automatic proof tools to prove some interesting properties about dimension(s) of feasible derivation trees of a given program.
\ \\

\end{abstract}

\input{intro}

\input{prelim}

\input{decomp}

\input{authorn}

\input{dimarg}

\input{related}

\input{experiments}

\input{conclusion}

\bibliographystyle{eptcs}
\bibliography{refs}
\end{document}

%% file: localdefs.tex
\newcommand{\integ}{{\sf int}}
\newcommand{\listint}{{\sf listint}}
\newcommand{\other}{{\sf other}}
\newcommand{\true}{\mathsf {true}}
\newcommand{\false}{\mathsf {false}}
\newcommand{\Bin}{{\sf Bin}}
\newcommand{\Dep}{{\sf Dep}}
\newcommand{\g}{{\sf g}}
\newcommand{\nong}{{\sf ng}}
\newcommand{\OL}{{\cal O}}
\newcommand{\M}{{\sf M}}
\newcommand{\R}{{\cal R}}
\newcommand{\A}{\mathcal{A}}

\newcommand{\body}{\mathcal{B}}
\newcommand{\B}{{\cal B}}
\newcommand{\C}{{\cal C}}
\newcommand{\D}{{\cal D}}
\newcommand{\X}{{\cal X}}
\newcommand{\V}{{\cal V}}
\newcommand{\Q}{{\cal Q}}
\newcommand{\F}{{\sf F}}
\newcommand{\N}{{\cal N}}
\newcommand{\Lang}{{\cal L}}
\newcommand{\powerset}{{\cal P}}
\newcommand{\FTA}{{\cal FT\!A}}
\newcommand{\Term}{{\sf Term}}
\newcommand{\Empty}{{\sf empty}}
\newcommand{\nonEmpty}{{\sf nonempty}}
\newcommand{\compl}{{\sf complement}}
\newcommand{\args}{{\sf args}}
\newcommand{\preds}{{\sf preds}}
\newcommand{\gnd}{{\sf gnd}}
\newcommand{\lfp}{{\sf lfp}}
\newcommand{\psharp}{P^{\sharp}}
\newcommand{\minimize}{{\sf minimize}}
\newcommand{\headterms}{\mathsf{headterms}}
\newcommand{\solvebody}{\mathsf{solvebody}}
\newcommand{\solve}{\mathsf{solve}}
\newcommand{\fail}{\mathsf{fail}}
\newcommand{\member}{\mathsf{memb}}
\newcommand{\ground}{\mathsf{ground}}

\newcommand{\raf}{{\sf raf}}
\newcommand{\qa}{{\sf qa}}
\newcommand{\spl}{{\sf split}}

\newcommand{\transitions}{\mathsf{transitions}}
\newcommand{\nonempty}{\mathsf{nonempty}}
\newcommand{\dom}{\mathsf{dom}}

\newcommand{\Args}{\mathsf{Args}}
\newcommand{\id}{\mathsf{id}}
\newcommand{\type}{\tau}
\newcommand{\restrict}{\mathsf{restrict}}
\newcommand{\any}{\top}
\newcommand{\dyn}{\top}
\newcommand{\dettypes}{{\sf dettypes}}
\newcommand{\Atom}{{\sf Atom}}

\newcommand{\chc}{{\sf chc}}
\newcommand{\deriv}{{\sf deriv}}

\newcommand{\vars}{\mathsf{vars}}
\newcommand{\Vars}{\mathsf{Vars}}
\newcommand{\range}{\mathsf{range}}
\newcommand{\varpos}{\mathsf{varpos}}
\newcommand{\varid}{\mathsf{varid}}
\newcommand{\argpos}{\mathsf{argpos}}
\newcommand{\elim}{\mathsf{elim}}
\newcommand{\pred}{\mathsf{pred}}
\newcommand{\predfuncs}{\mathsf{predfuncs}}
\newcommand{\project}{\mathsf{project}}
\newcommand{\reduce}{\mathsf{reduce}}
\newcommand{\positions}{\mathsf{positions}}
\newcommand{\contained}{\preceq}
\newcommand{\equivalent}{\cong}
\newcommand{\unify}{{\it unify}}
\newcommand{\Iff}{{\rm iff}}
\newcommand{\Where}{{\rm where}}
\newcommand{\State}{\mathsf{S}}
\newcommand{\qmap}{{\sf qmap}}
\newcommand{\fmap}{{\sf fmap}}
\newcommand{\ftable}{{\sf ftable}}
\newcommand{\Qmap}{{\sf Qmap}}
\newcommand{\states}{{\sf states}}
\newcommand{\head}{\tau}
\newcommand{\atomconstraints}{\mathsf{atomconstraints}}
\newcommand{\thresholds}{\mathsf{thresholds}}
\newcommand{\term}{\mathsf{Term}}
\newcommand{\trees}{\mathsf{trees}}
\newcommand{\renames}{\rho}
\newcommand{\renameps}{\rho_2}
\newcommand{\predicates}{\mathsf{Predicates}}
\newcommand{\query}{\mathsf{q}}
\newcommand{\ans}{\mathsf{a}}
\newcommand{\trace}{\mathsf{tr}}
\newcommand{\constr}{\mathsf{constr}}
\newcommand{\Iproj}{\mathsf{proj}}
\newcommand{\SAT}{\mathsf{SAT}}
\newcommand{\interpolant}{\mathsf{interpolant}}
\newcommand{\unknown}{?}
\newcommand{\rhs}{{\sf rhs}}
\newcommand{\lhs}{{\sf lhs}}
\newcommand{\unfold}{{\sf unfold}}
\newcommand{\arity}{{\sf ar}}
\newcommand{\AND}{{\sf AND}}

\newcommand{\atmost}[1]{\le #1}
\newcommand{\exactly}[1]{=#1}
\newcommand{\exceeds}[1]{>#1}
\newcommand{\anydim}[1]{\ge 0}

\def\ll{[\![}
\def\rr{]\!]}

\newcommand{\sset}[2]{\left\{~#1  \left|
                               \begin{array}{l}#2\end{array}
                          \right.     \right\}}

\newcommand{\qin}{\hspace*{0.15in}}
\newenvironment{SProg}
     {\begin{small}\begin{tt}\begin{tabular}[t]{l}}%
     {\end{tabular}\end{tt}\end{small}}
\def\anno#1{{\ooalign{\hfil\raise.07ex\hbox{\small{\rm #1}}\hfil%
        \crcr\mathhexbox20D}}}

\newtheorem{definition}{Definition}
\newtheorem{example}{Example}
\newtheorem{corollary}{Corollary}

\newtheorem{lemma}{Lemma}
\newtheorem{theorem}{Theorem}
\newtheorem{proposition}{Proposition}

%% file: intro.tex
\section{Introduction}

In this paper, we study the role of  \emph{tree dimension} in Horn clause analysis and verification.  The dimension of a tree is a
measure of its non-linearity -- for example a list of any length has dimension zero while a complete binary tree has dimension equal to its height. We apply this concept to trees corresponding to Horn clause derivations. A given set of Horn clauses $P$ can be transformed into a new set of clauses
$P^{\atmost{k}}$ (whose derivation trees are the subset of $P$'s derivation trees with dimension at most $k$) and  $P^{\exceeds{k}}$ (whose derivation trees have dimension at least $k+1$). Each such set of clauses represents an under-approximation of the original set of clauses and the proof for the original clauses can be constructed from their individual proofs. In order to prove some property of all derivations of $P$, we systematically
apply these transformations, for various values of $k$, to decompose the proof into separate proofs for $P^{\atmost{k}}$ and $P^{\exceeds{k}}$ (which could be
executed in parallel).

We prove each such set of clauses using abstract interpretation \cite{DBLP:conf/popl/CousotC77} over the domain of convex polyhedra \cite{Cousot-Halbwachs-78} as described in \cite{DBLP:conf/pepm/KafleG15}. Finally, the preliminary results in a set of Horn clause verification benchmarks show that this is a useful program transformation. This decomposition can also be viewed as refinement where one eliminates possibly infinite sets of program traces. As a result of this, the proof for the remaining part becomes simpler. 
To motivate  readers, we present an example set of constrained Horn clauses (CHCs) $P$ in Figure  \ref{exprogram} which defines the Fibonacci function. This is an interesting  problem whose dimension  depends on the input number  and its computations are trees rather than  linear sequences.
The main contributions of this paper are the following.
 \begin{figure}[t]
\centering
\begin{BVerbatim}
c1. fib(A, A):- A>=0,  A=<1.
c2. fib(A, B) :- A > 1, A2 = A - 2, fib(A2, B2),
           A1 = A - 1, fib(A1, B1), B = B1 + B2.
c3. false:- A>5, fib(A,B), B<A.          
\end{BVerbatim}
\caption{Example CHCs Fib:  it defines a Fibonacci function.}
 \label{exprogram}

\end{figure}

\begin{enumerate}

\item We describe how to generate  at-most k-dimension program and  at-least k-dimension program from a given program using the notion of tree dimension (Section \ref{prelim});

\item We give a verification algorithm for Horn clauses program based on its proof decomposition    (Section \ref{procverfication}); 

\item We give an alternative way  of generating  the at-least k-dimension program using the theory of finite tree automata (Section \ref{authorn});

\item  We demonstrate the feasibility of our approach in practice  applying it to non-linear Horn clause verification problems (Section \ref{experiments});

\item We instrument a program with its dimension and use  existing automatic verification tools to prove some interesting properties about its dimension  (Section \ref{proginstr}).

\end{enumerate}

%% file: prelim.tex
\section{Preliminaries}
\label{prelim}

A constrained Horn clause  is a first order  formula of the form 
$ p(X) \leftarrow   \C , p_1(X_1) , \ldots , p_k(X_k) $ ($k \ge 0$) (using Constraint Logic Programming (CLP) syntax),  where $\C$ is a conjunction of constraints with respect to some background theory, $X_i, X$  are (possibly empty) vectors of distinct variables, $p_1,\ldots,p_k, p$ are predicate symbols, $p(X)$ is the head of the clause and $\C , p_1(X_1) , \ldots , p_k(X_k)$ is the body.  A clause is called non-linear if it contains more than one atom in the body ($k > 1$), otherwise it is called linear.  A set of Horn clauses is sometimes called a program.

\begin{figure}[h!]
  \label{treedimension}
  \centering
    \includegraphics[width=0.70 \textwidth, height=45mm]{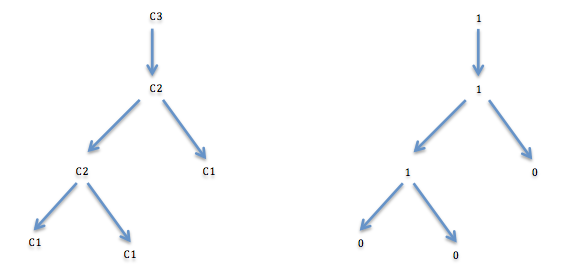}
    \caption{ (a) derivation tree of Fibonacci 3  and (b) its tree dimension.}
\end{figure}

A   labeled tree $c(t_1,\ldots,t_k)$  is a tree with its nodes labeled, where $c$ is a node label and $t_1,\ldots,t_k$ are labeled trees rooted at the children of the node and leaf nodes are denoted by $c$. 
\begin{definition}[Tree dimension (adapted from \cite{DBLP:conf/stacs/EsparzaKL07})]\label{treedim}
  Given a  labeled tree $t= c(t_1,\ldots,t_k)$, the tree dimension of $t$ represented as \(\mathit{dim}(t)\) is defined as follows: 

  \[
  \mathit{dim}(t)= \begin{cases}
  0 & \text{if } k=0 \\ 
  \max_{ i \in [1..k]} \mathit{dim}(t_i) &\text{if  there  is  a  unique  maximum}\\ 
  \max_{ i \in [1..k]} \mathit{dim}(t_i)+1 &\text{otherwise } 
\end{cases}
  \]

\end{definition}
Figure~\ref{treedimension} (a) shows a  derivation tree $t$ for  Fibonacci number  3 and  Figure~\ref{treedimension} (b) shows  its tree dimension.  It can be seen that $\mathit{dim}(t)=1$. This number is a measure of its non-linearity, the smaller the number the closer the tree is to a list.  Since it is not a perfect binary tree, the height of $t$ (3) is greater than its dimension.

 Given a set of CHCs $P$ and  $k \in \mathbb{N}$, we split each predicate $H$ occurring in $P$ into the predicates $H^ {\atmost{d}}$ and $ H^ {\exactly{d}}$ where $d \in \{ 0,1,\ldots,k\}$. Here $H^ {\atmost{d}}$ and $ H^ {\exactly{d}}$ generate trees of dimension at most $d$ and exactly $d$ respectively. 

\begin{definition}[ At-most-k-dimension program $P^{\atmost{k}}$]
\label{kdim-most}
 It consists of the following clauses (adapted  from \cite{DBLP:journals/corr/abs-1112-2864}):

\begin{enumerate}

\item Linear clauses:

 If $H  \leftarrow  \C \in P$ , then $H^ {\exactly{0}} \leftarrow  \C \in P^{\atmost{k}}$.
 
 If $H  \leftarrow  \C, B_1  \in P$  then $H^ {\exactly{d}} \leftarrow   \C,  B_1 ^{\exactly{d}} \in P^{\atmost{k}}$ for  $0 \le d \le k$.

\item Non-linear clauses: 

 If $H  \leftarrow  \C, B_1 , B_2 , \ldots,B_r  \in P$ with $r>1$:
\begin{itemize}
\item For $1 \le d \le k$, and $1 \le j \le r$:

Set  $Z_j =B_j^{\exactly{d}}$ and $Z_i = B_i ^{\atmost{d-1}}$ for $1 \le i \le r \wedge i \neq j$. Then: 
$H^ {\exactly{d}} \leftarrow   \C, Z_1,\ldots, Z_r \in P^{\atmost{k}}$.

\item 
 For  $1 \le d \le k$, and $J \subseteq \{1,\ldots, r\}$ with $\vert J \vert = 2$:
 
Set $Z_i =B_i^ {\exactly{d-1}}$ if $i \in J$ and $Z_i =B_i^ {\atmost{d-2}}$ if $i \in \{1,\ldots,r\} \setminus J$. If all $Z_i$ are defined, i.e., $d\geq2$ if 
$r > 2$, then:
$H^ {\exactly{d}} \leftarrow   \C, Z_1, \ldots, Z_{r} \in P^{\atmost{k}}$.

\end{itemize}

\item $\epsilon$-clauses: 

$H^ {\atmost{d}} \leftarrow H^ {\exactly{e}} \in P^{\atmost{k}}$ for $0 \le d \le k$ , and every $0 \le e \le d$.

\end{enumerate}
\end{definition}

 \begin{figure}[t]
 \centering
\begin{BVerbatim}
	%linear clauses
        1. fib(0)(A,A) :- A>=0, A=<1.
        2. false(0) :- A>5, B<A, fib(0)(A,B).
        %epsilon-clauses
        3. false[0] :- false(0).
        4. fib[0](A,B) :- fib(0)(A,B).
\end{BVerbatim}
\caption{Fib$^ {\atmost{0}}:$ at-most 0-dimension program of {Fib}.}
 \label{atmostk}
\end{figure}

The  at-most 0-dimension program  of  {Fib} in Figure \ref{exprogram} is depicted in Figure \ref{atmostk} (where 
the numbers on the first column  are not clause identifiers and are there for future reference).  In textual form we represent a predicate $p^{\atmost{k}}$ by \texttt{p[k]} and a predicate $p^{\exactly{k}}$ by \texttt{p(k)}.
Since some programs have derivation trees of unbounded dimension, trying to verify a property for its increasing dimension separately is not a practical strategy. To deal with this, we need some construction which characterises derivation trees of at-least $k$-dimension. Next we define this construction (\emph{at-least k-dimension program}). For this, we split each predicate $H$ occurring in $P$ into the predicates $H^ {\exceeds{d}}$ and $H^ {\anydim{d}}$ where $d \in \{ 0,1\ldots,k\}$. Here $H^ {\exceeds{d}}$  generates trees of dimension at-least $d+1$ and $H^ {\anydim{d}}$  generates trees of any dimension.

\begin{definition}[At-least k+1-dimension program $P^ {\exceeds{k}}$] 
In \hspace{-0.5pt}addition \hspace{-0.5pt}to \hspace{-0.5pt}the \hspace{-0.5pt}linear, \hspace{-0.5pt}non-linear \hspace{-0.5pt}and \hspace{-0.5pt}$\epsilon$-clauses 
from Definition \ref{kdim-most} (with each predicate $H^{\atmost{k}}$ and $H^ {\exactly{k}}$ from $P^{\atmost{k}}$  renamed to $H^{\exceeds{k}}$   and $H^ {\anydim{k}}$ respectively), the at-least k+1-dimension program $P^ {\exceeds{k}}$ consists of the following clauses:
\label{kdim-least}

\begin{enumerate}

\item Link-clauses: 

For each $H  \leftarrow  B \in P$  there is a clause $ H^ {\anydim{0}} \leftarrow H \in P^{\exceeds{k}}$.

\item Original clauses:

All clauses in  $P$ are also in $P^{\exceeds{k}}$.

\end{enumerate}

\end{definition}

The  at-least 1-dimension program   of Fib in Figure \ref{exprogram} is depicted in Figure  \ref{atleastk}.  In textual form we represent a predicate $p^{\exceeds{k}}$ by \texttt{p\{k\}} and a predicate $p^{\anydim{k}}$ by \texttt{p<k>}.
 
 \begin{figure}[t]
 \centering
\begin{BVerbatim}
%linear clauses
fib<0>(A,A) :- A>=0, A=<1.
false<0> :- A>5, B<A,fib<0>(A,B).
%epsilon-clauses
false{0} :- false<0>.
fib{0}(A,B) :- fib<0>(A,B).
%link clauses
false<0> :- false.
fib<0>(A,B) :- fib(A,B).
%original clauses (all clauses)
fib(A, A):- A>=0,  A=<1.
false:- A>5, fib(A,B), B<A.
fib(A, B) :- A > 1, A2 = A - 2, fib(A2, B2),
           A1 = A - 1, fib(A1, B1), B = B1 + B2.
\end{BVerbatim}
\caption{Fib$^ {\exceeds{0}}:$ at-least 1-dimension program of Fib.}
 \label{atleastk}
\end{figure}

%% file: decomp.tex
\section{Procedure for verification}
\label{procverfication}

Given a set of CHCs $P$ (including clauses with $\false$ head, also known as \emph{integrity constraints}), the CHC verification problem is to check whether there exists a model of $P$. This is equivalent to checking whether there is any feasible derivation tree for $\false$; if there is such a derivation then there is no model. We say $P$ is safe if it has a model and unsafe if it has no model. The procedure VERIFY($P$) is described in algorithm \ref{alg:verify}. VERIFY makes use of the procedure SAFE($P$) in the Algorithm \ref{alg:verify}, which is an oracle that returns \emph{safe}, \emph{unsafe} or \emph{unknown}.  The oracle is sound: if SAFE($P$) returns \emph{safe}  (\emph{unsafe}) then $P$ is safe (unsafe).
SAFE could be any existing automatic Horn clause solver  \cite{DBLP:conf/tacas/GrebenshchikovGLPR12,DBLP:conf/vmcai/KafleG15, DBLP:conf/pepm/KafleG15,DBLP:conf/sat/HoderB12,DBLP:conf/tacas/AngelisFPP14}.  When it cannot verify a program within a given time limit, the \emph{unknown} answer is emitted. A given set of Horn clauses $P$ can be transformed into a new set of clauses $P^{\atmost{k}}$ and $P^{\exceeds{k}}$. In order to prove some property of all derivations of $P$, we systematically apply these transformations, for various values of $k$, to decompose the proof into separate proofs for $P^{\atmost{k}}$ \emph{(line 4)} and $P^{\exceeds{k}}$ \emph{(line 9)}. If both are safe then $P$ is \emph{safe}. If one of them is unsafe then $P$ is \emph{unsafe}. If an oracle cannot prove whether $P^{\atmost{k}}$ is \emph{safe/unsafe} then we return an \emph{unknown} answer (we assume that the oracle would also return  \emph{unknown} for larger values of $k$). But if it cannot prove whether $P^{\exceeds{k}}$ is \emph{safe/unsafe} then we try the {\it while loop } in the algorithm \ref{alg:verify} with $k=k+1$.   

\begin{figure}[t]
\centering
\RestyleAlgo{boxruled}
\LinesNumbered
\begin{algorithm}[H]{Procedure VERIFY ($P$) }

\label{alg:verify}
\caption{Verification  algorithm for Horn clauses}
\KwIn{Set of CHCs $P$}
\KwOut{\emph{safe, unsafe, unknown}}
  initialization: $k \gets 0$ \\
  \While {true} { 
   generate $P^{\atmost{k}}$ \\
   $r_1 \gets$ SAFE($P^{\atmost{k}}$) \\
  \If {$r_1 \neq$ safe} {\Return $r_1$} 
    generate $P^{\{k\}}$  \\
  $r_1' \gets$ SAFE( $P^{\exceeds{k}}$) \\
  \If {$r_1' \neq$ unknown} {\Return $r_1'$} 
 $k \gets k+1$ 
 }
 \vskip 2mm

  \end{algorithm}
  \end{figure}

 One possible optimisation that we can make in Algorithm \ref{alg:verify} is to consider  $P^{\exceeds{k}}$ instead of $P$ in the next iteration of the {\it while loop} if we reach \emph{line 14}. This is because at this stage we have already proven the safety of  $P^{\atmost{k}}$.

The soundness of Algorithm \ref{alg:verify} is captured by the following lemma and proposition. 

\begin{lemma}[Decomposition by dimension] 
For all $k$, program $P$ is safe if and only if both $P^{\atmost{k}}$ and $P^{\exceeds{k}}$ are safe.
\end{lemma}

\begin{proposition}[Soundness] 
  If Algorithm \ref{alg:verify} returns  \emph{safe} then the input program is safe. If it returns \emph{unsafe} then the program is unsafe.
\end{proposition}

%% file: authorn.tex
\section{Dimension decomposition using finite tree automata}
\label{authorn}

In this section, we show an alternative method for constructing an at-least k-dimension program, using operations on finite tree automata (FTAs). We first describe the connection between Horn clauses and FTAs and show how to construct an FTA from a set of Horn clauses. 

\subsection{Trace automata for CHCs}
We add identifiers to clauses, whose purpose is to act as constructors of trace trees representing derivations. The identifiers are chosen from a 
set $\Sigma$ of ranked function symbols.  If $P$ is a set of CHCs, let $\id_P: P \rightarrow \Sigma$ be an assignment of function symbols to clauses, such that for every clause $cl \in P$, the arity of $\id_P(cl)$ equals the number of atoms in the body of $cl$. We allow the same symbol to be assigned by $\id_P$ to more than one clause. We can also identify the predicates whose derivations are of interest (the \emph{accepting} predicates in Definition \ref{trace-fta}).
\begin{definition}[Trace FTA for a set of CHCs] 
\label{trace-fta} 
Let $P$ be a set of CHCs, $\Sigma$ be a set of ranked function symbols and $\id_P: P \rightarrow \Sigma$ be a mapping from clauses to function symbols of appropriate arity.  Let $F$ be a set of predicates from $P$ called the \emph{accepting} predicates. 
Define the trace FTA for $P$ as $\A_P^F = (Q,F,\Sigma,\Delta)$ where
\begin{itemize}
\item
$Q$ is the set of predicate symbols of $P$;
\item
$F \subseteq Q$ is the set of accepting predicate symbols;
\item
$\Sigma$ is a set of function symbols;
\item
$\Delta = \{c(p_1,\ldots,p_k) \rightarrow p \mid  cl \in P,   ~cl = p(X) \leftarrow \C, p_1(X_1), \ldots, p_k(X_k),   ~c = \id_P(cl) \}$.
\end{itemize}
If  $F$ is the set of all predicate symbols occurring in the clauses we omit the superscript $F$ from $\A_P^F$.
\end{definition}

The  set of trees accepted by $\A_P^F$ is written $\Lang(\A_P^F)$.  Elements of $\Lang(\A_P^F)$ are called the trace trees for $P$.
$\Lang(\A_P^F)$ is isomorphic to the set of (successful and unsuccessful) derivation trees (for atomic formulas with accepting predicates) constructible from $P$ and from now on we identify trace trees with derivations. We do not define derivation trees formally here, but refer to the notion of an AND-tree in the literature \cite{Stark89,Gallagher-Lafave-Dagstuhl}.  

\begin{example}
\label{exautP}
Let $P$ be the set of CHCs in Figure \ref{exprogram} and let $F=\{\mathtt{fib}, \mathtt{false}\}$. Let $\id_P$ map the clauses to $c_1, c_2, c_3$ respectively. Then $\A_{P}^F = (Q,F,\Sigma,\Delta)$ where:
\[
\begin{array}{llllll}
Q &=& \{\mathtt{fib}, \mathtt{false}\}~~~&\Delta &=& \{ c_1 \rightarrow \mathtt{fib},\\
 \Sigma &=& \{ c_1, c_2,  c_3\} &&& ~c_2(\mathtt{fib}, \mathtt{fib}) \rightarrow \mathtt{fib}, \\
& &&	&&~c_3( \mathtt{fib}) \rightarrow \mathtt{false}  \}
\end{array}
\]
Figure \ref{treedimension}(a) shows a trace tree recognised by this FTA. The tree can also be written $c_3(c_2(c_2(c_1,c_1),c_1))$.
\end{example}

If a mapping $\id_P: P \rightarrow \Sigma$ assigns a unique identifier to each clause, that is, $\id_P$ is injective, then there is an inverse mapping $\id^{-1}: \range(\id_P) \rightarrow P$.  

\begin{definition}[$\chc_{\id}(\A)$]\label{fta2chc}
Given an FTA $\A = (Q,F,\Sigma,\Delta)$ and an injective mapping $\id$ such that $\Sigma \subseteq \range(\id)$, we can construct a set of CHCs from $\A$, called $\chc_{\id}(\A)$, defined as follows:
\[
\begin{array}{ll}
\chc_{\id}(\A) = \{q(X) \leftarrow \C, q_1(X_1),\ldots,q_n(X_n) \mid &c(q_1,\ldots,q_n) \rightarrow q \in \Delta, \\
&										\id^{-1}(c) = q(X) \leftarrow \C, q_1(X_1),\ldots,q_n(X_n)\}
\end{array}
\]
The set of accepting predicates of $\chc_{\id}(\A)$ is defined to be $F$.
\end{definition}
\noindent
In the definitions we reuse the states in the FTA as predicate symbols in the constructed clauses.  In practice we use some injective renaming function from states to predicates in the constructed program.
Further discussion of the mappings between CHCs and FTAs can be found in \cite{DBLP:conf/vmcai/KafleG15}.   By construction, the derivations of $\chc_{\id}(\A)$ (for the accepting predicates) correspond to the elements of 
$\Lang(\A)$.

\subsection{Construction of the at-least k-dimensional program using FTA operations}

In the construction of the at-least k-dimension program $P^{\exceeds{k}}$ in Definition \ref{atleastk}, the original program clauses from $P$ are included in the generated clauses. The presence of the original clauses suggests that the ``decomposed" verification problem for $P^{\exceeds{k}}$ is as hard as the original problem for $P$, since it contains the clauses of $P$ as well as others, and so this form might not lend itself to verification.


Thus in the following construction we build $P^{\exceeds{k}}$ based on FTA language difference, and the original clauses are not copied to the at-least k-dimension program.  We first define a general FTA-difference for CHCs.

\begin{definition}[FTA-difference for CHCs]\label{fta_diff}
Let $P$ and $Q$ be sets of CHCs,  $F_1$ and $F_2$  their respective accepting predicates and  $\id_P:P \rightarrow \Sigma$ and $\id_Q: Q\rightarrow \Sigma$  their respective identifier assignments, where $\id_P$ is injective. Let $\A_P^{F_1}$ and $\A_Q^{F_2}$ be the trace FTAs constructed from $P, Q$ respectively.  Then the FTA-difference of $P$ and $Q$ (with their respective accepting predicates) written $P^{F_1}-Q^{F_2}$, is given as $\chc_{\id_P}(\A_P^{F_1} \setminus \A_Q^{F_2})$ where $\setminus$ is the difference of FTAs \cite{Comon}.  The set of accepting predicates is the set of accepting states for the difference FTA.
\end{definition}
The set of derivations for $P^{F_1}-Q^{F_2}$ contains, by construction, those derivations of $P^{F_1}$ that are not derivations of $Q^{F_2}$.
We now apply these notions to the verification procedure based on decomposition. We are given a set of CHCs $P$, with accepting predicates $F = \{\false\}$. In the program $P^{\atmost{k}}$, the set of accepting predicates is $F^k = \{\false^{\atmost{k}}\}$. Note that  we can ignore the derivations for the other predicates of the form $\false^{\atmost{j}}$ or $\false^{\exactly{j}}$  since $\false^{\atmost{k}}$ by construction accumulates  their derivations,  for all $j \le k$.

\subsubsection{Assignment of identifiers in the at-most-k-dimension program}

Given a program $P$ and the at-most-k-dimension program $P^{\atmost{k}}$, we intend to construct the difference
$P^{\{\false\}} - P^{\atmost{k} \{\false\}}$ using Definition \ref{fta_diff}.
In order to do so, we first need to construct the identifier assignment $\id_{ P^{\atmost{k} }}$ so as to preserve trace trees from $P$. 
This requires the modification of $P^{\atmost{k}}$ to eliminate the $\epsilon$-clauses, as follows.
\begin{definition}[Unfolding of $\epsilon$-clauses in $P^{\atmost{k}}$]\label{eps_unfold}

Let\hfill $P^{\atmost{k}}$\hfill be\hfill the at-most-k-dimension\hfill program\hfill obtained\\ from 
$P$ using Definition \ref{kdim-most}.  Replace each $\epsilon$-clause of form $H^{\atmost{d}} \leftarrow H^{\exactly{e}}$ by the set of clauses $H^{\atmost{d}} \leftarrow B$, where $H^{\exactly{e}} \leftarrow B$ is either a linear or non-linear clause in $P^{\atmost{k}}$.
\end{definition}
The elimination of $\epsilon$-clauses is an instance of the well-known unfolding transformation which preserves the derivability of atomic formulas.  In other words an atom $A$ is derivable from a program $P$ if and only if it is derivable after applying the unfolding transformation \cite{Pettorossi-Proietti}.

In the following definition, the clause identifiers are chosen for clauses in $P^{\atmost{k}}$  Informally, every clause of $P^{\atmost{k}}$  inherits the clause identifier for the clause in $P$ from which it originates.  More precisely we define the clause identifiers for $P^{\atmost{k}}$ as follows.

\begin{definition}[Assignment of clause identifiers in $P^{\atmost{k}}$]\label{id_pk}

Let $P^{\atmost{k}}$ be the at-most-k-dimension program obtained from $P$ using Definition \ref{kdim-most}, with $\epsilon$-clauses eliminated according to Definition \ref{eps_unfold}.
Each clause of $P^{\atmost{k}}$ is a linear, non-linear or an $\epsilon$-unfolded-clause.  The clause identifiers are assigned in two steps as follows.

\begin{enumerate}
\item
Assign to each linear or non-linear clause the clause identifier from the clause in $P$ from which it is derived in Definition \ref{kdim-most}.
\item
Assign to each  unfolded $\epsilon$-clause the clause identifier for the linear or non-linear clause used to unfold it using Definition \ref{eps_unfold}.
\end{enumerate}

\end{definition}
We are now in a position to compare the sets of trace trees for $P$ and $P^{\atmost{k}}$ using their respective FTAs.
\begin{lemma}
Let $P$ be a set of CHCs and let $\id_P: P \rightarrow \Sigma$ be an injective function assigning clause identifiers to $P$. Let $F_1 = \{\false\}$. 
Let $k\ge 0$ and let $P^{\atmost{k}}$ be the at-most-k-dimension program obtained from $P$ using Definition \ref{kdim-most} with $\epsilon$-clauses unfolded using Definition \ref{eps_unfold} and let $F_2 =\{\false^{\atmost{k}}\}$.  Then $\Lang(\A_{P^{\atmost{k}}}^{F_2}) = \{t \mid t \in \Lang(A_P^{F_1}), \mathit{dim}(t) \le k\}$.
\end{lemma}
The proof is by induction on derivations in $P^{\atmost{k}}$ and uses the correspondence of the clause identifiers as set up in Definition \ref{id_pk}.

\begin{theorem}
Let $P$ be a set of CHCs and let $\id_P: P \rightarrow \Sigma$ be an injective function assigning clause identifiers to $P$.
Let $k\ge 0$ and let $P^{\atmost{k}}$ be the at-most-k-dimension program obtained from $P$ using Definition \ref{kdim-most} with $\epsilon$-clauses unfolded using Definition \ref{eps_unfold}. Then $\false$ is derivable from $P-P^{\atmost{k}}$ if and only if $\false^{\exceeds{k}}$ is derivable from $P^{\exceeds{k}}$.
\end{theorem}
Thus we have shown a different method of constructing the at-least k-dimension program $P^{\exceeds{k}}$, namely by taking the difference of $P$ with $P^{\atmost{k}}$, which contains only derivations (for its accepting predicates) that have dimension greater than $k$.

Details on difference construction can be found in \cite{DBLP:conf/vmcai/KafleG15}. We construct the difference of two FTAs by (1) standardising apart the predicate names; (2) forming the union of the two FTAs; (3) determinising the union; (4) removing from the determinised FTA all states (and transitions that contain them) that contain an accepting state of the second FTA.  Note that the set of states of the determinised FTA is a subset of the powerset of the original states.  Note that determinisation of FTAs is often considered prohibitively complex even for small FTAs. We use a recent optimised FTA determinisation algorithm \cite{TR145RUC},
returning a compact form of the determinised called product form, which can be used directly in constructing the resulting clauses.

\begin{example}
We illustrate this through an example using Fib$^{\atmost{0}}$ (Figure \ref{atmostk}). The clauses 1 and 2 in Fib$^{\atmost{0}}$, will have $c_1$ and $c_3$ as identifiers since they were derived respectively from the clauses $c_1$ and $c_3$ in Fib (Figure \ref{exprogram}). By unfolding $\epsilon$-clauses (clauses 3 and 4) using respectively clauses 2 and 1 in Figure \ref{atmostk}, we obtain \emph{\texttt{false[0] :- A>5, B<A, fib(0)(A,B) and  fib[0](A,B) :-A>=0, A=<1}}. They will have identifiers $c_3$ and $c_1$ respectively. Therefore, the clauses in Fib$^{\atmost{0}}$ will have the  identifiers assigned as shown in Figure \ref{fib0unfolded}. 

\end{example}
\begin{figure}[t]

\centering
\begin{BVerbatim}
c1. fib(0)(A,A) :- A>=0, A=<1.
c3. false(0) :- A>5, B<A, fib(0)(A,B).
c3. false[0] :- A>5, B<A, fib(0)(A,B).
c1. fib[0](A,B) :-A>=0, A=<1.
\end{BVerbatim}

\caption{Fib$^{\atmost{0}}$ after unfolding $\epsilon$-clauses and assigning clause identifiers.}
\label{fib0unfolded}
\end{figure}

After assigning  identifiers to each of the clauses in Fib$^{\atmost{0}}$, we can construct an FTA  corresponding to it using Definition \ref{trace-fta}, and obtain the FTA shown in Figure \ref{fib0-fta}: as before we represent a predicate $p^{\atmost{k}}$ by \texttt{p[k]} and a predicate $p^{\exactly{k}}$ by \texttt{p(k)}.

\begin{figure}[h]

\[
\begin{array}{llllll}
Q = \{\texttt{fib(0)}, \texttt{  false(0)}, \texttt{ false[0]},   \texttt{ fib[0]} \}~~~&\Delta = \{ c_1 \rightarrow \texttt{ fib(0)},\\
F = \{ \texttt{false[0]} \} & ~c_3( \texttt{ fib(0)}) \rightarrow \texttt{ false(0)}, \\
 \Sigma = \{ c_1, c_3\} & 	~ c_3(\texttt{ fib(0)}) \rightarrow \texttt{ false[0]} , \\
 &    c_1 \rightarrow \texttt{ fib[0]} \}
\end{array}
\]

\caption{FTA \((Q,F,\Sigma,\Delta)\) corresponding to Fib$^{\atmost{0}}$.}
\label{fib0-fta}
\end{figure}


The difference FTA between  $\A_{Fib}^{\{\false\}}$  and  $\A_{Fib^{\atmost{0}}}^{\{\false^{\atmost{0}}\}}$   accepts trees rooted at  $\false$ which have dimension greater than 0. The determinised FTA (DFTA) constructed as explained above  is shown in the Figure \ref{fib-fib0.dfta}.   DFTA states are sets of predicates, and we represent a set using square brackets instead of curly brackets in the code, e.g. \texttt{[fib(0), fib[0], fib]}.  Furthermore the product form referred to above contains set of DFTA states, such as 
\texttt{[[fib(0), fib[0], fib], [fib]]}.

\begin{figure}[h]

\centering
\begin{BVerbatim}
c1 -> [fib(0), fib[0], fib].
c2([[fib(0), fib[0], fib], [fib]],
	              [[fib(0), fib[0], fib], [fib]]) -> [fib].
c3([[fib]]) -> [false].
\end{BVerbatim}

\caption{Transitions of the determinised FTA.}
\label{fib-fib0.dfta}
\end{figure}

We can generate a new program  
from this DFTA together with the original program Fib following the approach taken in \cite{DBLP:conf/vmcai/KafleG15} obtaining the  program in Figure \ref{atleastkfib}. It should be noted that the  derivation trees rooted at $\false$ have dimension at-least 1. Now verification of the original program Fib is decomposed into 
verifying the program in Figure \ref{atmostk} (where \texttt{false[0]} is replaced by \texttt{false} and the program in Figure \ref{atleastkfib}. 

\begin{figure}[h]
\centering
\begin{BVerbatim}
fib_0(A,A) :- A>=0, A=<1.
fib(A,B) :-  A>1,C=A-2,D=A-1,B=E+F,fib_1(C,F), fib_1(D,E).
false :- A>5, B<A, fib(A,B).
fib_1(A,B) :- fib_0(A,B).
fib_1(A,B) :-  fib(A,B).
\end{BVerbatim}
\caption{ At-least 1-dimension program of Fib produced using the difference of FTAs}
 \label{atleastkfib}
\end{figure}

%% file: dimarg.tex
\section{Program instrumentation with dimension}
\label{proginstr}

The dimension of successful derivations in a set of CHCs is not always obvious from the text of the clauses. In some cases a bound on the dimension is clear from the form of the clauses; for instance all derivations using a set of linear clauses clearly have dimension zero. But consider the well known 91-function of McCarthy\footnote{http://en.wikipedia.org/wiki/McCarthy\_91\_function}, represented in Figure \ref{mc91program} using Horn clauses.

 \begin{figure}
 \centering
\begin{BVerbatim}
mc91(N,X) :-  N > 100,  X = N-10.
mc91(N,X) :-  N =< 100,  Y = N+11,
              mc91(Y,Y2), mc91(Y2,X).
\end{BVerbatim}
\caption{ McCarthy's 91-function defined as Horn clauses}
 \label{mc91program}
\end{figure}
Although it is possible to construct derivation trees of arbitrary dimension using the clauses in Figure \ref{mc91program}, the dependencies between the two recursive calls to \texttt{mc91} imply that no \emph{successful} derivation has dimension greater than 2. We now show how to establish this using a transformation to instrument the clauses with dimension information, and then use automatic verification tools to establish properties of the dimension.

\begin{definition}[Dimension-instrumented clauses]\label{dim_instr}
Let $P$ be a set of CHCs.  Define the set $P_{dim}$ of CHC as follows.
\begin{itemize}
\item
For each predicate $p$ of arity $m$ define a predicate $p'$ of arity $m+1$.
\item
For each clause in $P$ of the form
$$p(X) \leftarrow \C, p_1(X_1),\ldots,p_n(X_n)$$
construct a clause 
$$p'(X,K) \leftarrow \C, p_1^{'}(X_1,K_1),\ldots,p_n^{'}(X_n,K_n), dim_n(K_1,\ldots,K_n,K)$$
in $P_{dim}$, where $K_1,\ldots,K_n,K$ are variables added as the final argument for their respective predicates, and $dim_n(K_1,\ldots,K_n,K)$ is defined according to the rules in Definition \ref{treedim} for determining the dimension of a tree.
\end{itemize}

\end{definition}

\begin{example}
The dimension-instrumented version of the McCarthy 91-function contains the following clauses.
\begin{verbatim}
mc91(N,X,K) :-  N > 100,  X = N-10, dim0(K).
mc91(N,X,K) :-  N =< 100,  Y = N+11,
     mc91(Y,Y2,K1), mc91(Y2,X,K2), dim2(K1,K2,K).
dim0(K):-K=0.
dim2(K1, K2, K3):-K1>=K2+1, K3=K1.  
dim2(K1, K2, K3):-K2>=K1+1, K3=K2.    
dim2(K1, K2, K3):- K1=K2, K3 = K1+1.
\end{verbatim}
\end{example}
Using the instrumented program we can try to prove information about the dimension, such as upper or lower bounds or other relationships between the dimension and other predicate arguments. 
It follows from the undecidability result of Gruska~\cite{Gruska71b} on
context-free grammars, that the problem of determining whether the dimension of
set of CHC is bounded by a constant is, in general, undecidable. 

\begin{example}
To establish that the upper bound of successful derivations is 2, for facts \emph{\texttt{mc91(X,Y)}}, we add the following integrity constraint to the dimension-instrumented clauses.
\begin{verbatim}
false :- K > 2, mc91(X,Y,K).
\end{verbatim}
The clauses together with the integrity constraint are given to an automatic solver for Horn clauses \cite{DBLP:conf/tacas/GrebenshchikovGLPR12,DBLP:conf/vmcai/KafleG15}, which are able to prove the safety of the clauses and thus establish the upper bound of 2.
\end{example}

In the next example, we show that the dimension can depend on the values of other predicate arguments.  
\begin{example}
The dimension-instrumented version of the \emph{Fib} clauses is shown in Figure \ref{inprogram}.  The property to be proved is that the dimension of \emph{Fib} is lesser or equal to the  half of its input value, expressed by the integrity constraint \emph{\texttt{false:- fib(A,B, K), 2*K -1>=A}}. Again, this property is established by applying a Horn clause solver to prove the safety of the clauses together with the integrity constraint.
\end{example}

 \begin{figure}
 \centering
\begin{BVerbatim}
fib(A, A, K):-  A>=0,  A=<1, dim0(K).
fib(A, B, K) :- A > 1, A2 = A - 2, fib(A2, B2, K1),
           A1 = A - 1, fib(A1, B1, K2), B = B1 + B2, dim2(K1, K2, K).
dim0(K):-K=0.           
dim2(K1, K2, K3):-K1>=K2+1, K3=K1.  
dim2(K1, K2, K3):-K2>=K1+1, K3=K2.    
dim2(K1, K2, K3):- K1=K2, K3 = K1+1.           
\end{BVerbatim}
\caption{Fib program instrumented with its dimension}
 \label{inprogram}
\end{figure}


\begin{example}
We present the well known counting change example taken from \cite[Chapter 1]{DBLP:books/mit/AbelsonS96}. The Figure \ref{counting-change} shows its CLP encoding and the Figure \ref{counting-change-dim} shows the dimension-instrumented version in CLP. The property of interest is to relate the number of different coins (counts) with the program dimension. We can establish that the dimension is at most the number of different coins as expressed by the integrity constraint \emph{\texttt{false :- B>=1, K > B, cc(A, B, C, K)}}. 

\end{example}

 \begin{figure}
 \centering
\begin{BVerbatim}
% base case: that is a hit
cc(0, Y, 1) :- Y>0.
% base case: that is a miss
cc(X, _, 0) :- X<0.
cc(_, Y, 0) :- Y=<0.
%inductive case
cc(X, Y, Z) :- X>0, kinds_of_coins(Y,A), 
                    X1 = X-A, cc(X1, Y, Z1), 
                    Y1 = Y-1, cc(X, Y1, Z2), Z = Z1 + Z2.
kinds_of_coins(A,B) :- A >= 1, B >= 1.
        
\end{BVerbatim}
\caption{Counting change example encoded as CLP clauses}
 \label{counting-change}
\end{figure}

 \begin{figure}[h!]
 \centering
\begin{BVerbatim}
cc(0, Y, 1,K) :- Y>0, dim0(K).
cc(X, _, 0,K) :- X<0, dim0(K).
cc(_, Y, 0,K) :- Y=<0, dim0(K).
cc(X, Y, Z,K) :- 
	X>0, kinds_of_coins(Y,A, K0), X1 = X-A, 
	cc(X1, Y, Z1,K1), Y1 = Y-1, cc(X, Y1, Z2,K2), 
	Z = Z1 + Z2, dim3(K0, K1,K2,K).
kinds_of_coins(A,B, K) :- A >= 1, B >= 1, dim0(K).	
dim3(K0, K1,K2,K):-
	dim2(K0, K1, K3), dim2(K3,K2, K).
%predicates dim0(K) and dim2(K1, K2, K) are defined as above 	
        
\end{BVerbatim}
\caption{Counting change example instrumented with its dimension}
 \label{counting-change-dim}
\end{figure}

In general, verifying whether a program has a certain dimension is as challenging as proving any other properties of the program. But in some cases the knowledge of program dimension is useful for proving other program properties.  For instance, using the knowledge that the McCarthy 91-function  has dimension at most 2 would allow us to restrict the proof of any program property relating to successful derivations to the program $P^{[2]}$ where $P$ is the set of clauses for the McCarthy 91-function.

%% file: related.tex
\section{Related Work}
\label{rel}

The notion of dimension of a tree has a long history in science (starting with
Geology) which has been detailed by Esparza \textit{et al.}
\cite{esparzalata2014}. However, the use of dimension for program verification is
more recent.  Ganty and Iosif used it \cite{tacas13} for computing summaries of
programs with procedures whose variables (global, local and parameters) take
their value from the set of integers. Roughly speaking, the method they define
first computes procedure summaries for all  derivation trees of dimension \(0\), then they compute summaries for derivation trees of dimension \(1\) reusing the summaries computed for dimension \(0\)
and so on.

Decomposition can be compared to refinement techniques based on automata \cite{DBLP:conf/sas/HeizmannHP09,DBLP:conf/cav/HeizmannHP13,DBLP:conf/vmcai/KafleG15} in which the aim is to eliminate sets of program traces that have been shown to be safe.  Proof of the safety of a given dimension or dimensions of a set of clauses allows those dimensions to be eliminated, focusing the proof on the remaining dimensions.  Our decomposition technique offers a very precise and practical approach to checking and eliminating infinite sets of traces.

%% file: experiments.tex
\section{Experimental results}
\label{experiments}
We carried out an experiment on a set of 16 non-linear CHC verification problems taken from  the repository\footnote{https://svn.sosy-lab.org/software/sv-benchmarks/trunk/clauses/LIA/Eldarica/RECUR/} of software verification benchmarks. Our aim in the current paper is not to make a systematic comparison with other verification techniques;  these are exploratory experiments to establish whether dimension-based decomposition is practical. The results are summarized in Table~\ref{tbl:exp}. 
\begin{wraptable}{r}{7.1cm}
\caption{Experimental  results on non-linear CHC verification problems}\label{tbl:exp}
    \begin{tabular}{|l|l|l|l|}
    \hline
    {\bf Program}       & {\bf Result} & {\bf Time(s)} & {\bf dim(k)} \\ \hline
    addition                       & safe                    & 4                     & 0                       \\ \hline
    bfprt                          & safe                    & 4                     & 0                       \\ \hline
    binarysearch                   & safe                    & 4                     & 0                       \\ \hline
    countZero                      & safe                    & 3                     & 0                       \\ \hline
    floodfill                      & safe                    & 3                     & 0                       \\ \hline
    identity                       & safe                    & 4                     & 0                       \\ \hline
    merge                          & safe                    & 5                     & 0                       \\ \hline
    palindrome                     & safe                    & 3                     & 0                       \\ \hline
    fib                            & safe                    & 4                     & 0                       \\ \hline
    mc91                           & safe                    & 4                     & 0                       \\ \hline
    revlen                         & safe                    & 4                     & 0                       \\ \hline
    running                        & unsafe                  & 6                     & 1                       \\ \hline
    triple                         & unsafe                  & -                     & -                       \\ \hline
    buildheap                      & unsafe                  & -                     & -                       \\ \hline
    parity                         & unsafe                  & 4                     & 0                       \\ \hline
    remainder                      & unsafe                  & 4                     & 0                       \\ \hline
    {\bf avg. time(s)}  & ~                       & 4                     & ~                       \\ \hline
    \end{tabular}
\end{wraptable} 
Columns {\bf Program}, {\bf Result}, {\bf Time} and {\bf dim(k)} respectively represent a program, its verification result using our  approach, time in seconds taken to generate the programs and solve it and a value of a proof decomposition parameter $k$. 

For the  safety check (the procedure \emph{SAFE} in Algorithm \ref{alg:verify})  we use the verification procedure described in \cite{DBLP:conf/pepm/KafleG15} which uses abstract interpretation over the domain of convex polyhedra, with a timeout of 5 minutes. The symbol ``-'' in Table~\ref{tbl:exp} denotes that we were unable to solve these problems within the given time. 
Our  approach solves 14 out of 16  problems with an average time of 4 seconds (over the solved problems). Our previous approach based on refinement with finite tree automata described  in \cite{DBLP:conf/vmcai/KafleG15}  solves 1 more additional problem, that is, {\it triple} than our current approach. These examples were also run on QARMC \cite{DBLP:conf/pldi/GrebenshchikovLPR12}  which solves all the problems (much faster).

Most of the problems are solved when we decompose the proof with the value of  $k=0$. This indicates that separating the proofs for linear programs eases the verification task. The splitting induced as a result of  separating a set of traces has an effect on delaying join and widening operations during convex polyhedra analysis which increases its precision. In addition to this, some of the case base proofs (for example conditionals)  becomes a normal proof without conditionals due to proof separation and the process of finding invariants becomes easier.

%% file: conclusion.tex
\section{Conclusion and future work}

We presented a program transformation approach to Horn clause verification using the notion of \emph{tree
dimension} to decompose the verification problem by separating dimensions.  We presented one algorithm based on this idea which yielded preliminary results on set of non-linear Horn clause verification benchmarks, showing that the approach is feasible and this transformation is useful both for proving safety of a program as well as for finding bugs. 

Other ideas of program verification based on tree-dimension are worth investigating, including proof by induction based on tree dimension,  and further investigation of proof strategies that could exploit knowledge of dimension bounds (such as those discussed in Section \ref{proginstr}).

Although it is formulated in the context of Datalog, it is known from Afrati
\textit{et al.}~\cite{DBLP:journals/tcs/AfratiGT03} that a set of CHC of bounded
dimension can be turned into an equivalent set of linear CHC. The exact
complexity of their procedure is still open.
